\documentclass{proclund}

\newcommand{\Oxg}{\mbox{${}^{16}{\rm O}$}}
\newcommand{\Carb}{\mbox{${}^{14}{\rm C}$}}

\newcommand{\Oeep}{\mbox{$^{16}$O(e,e$'$p)}}
\newcommand{\Oeepp}{\mbox{$^{16}$O(e,e$'$pp)}}

\newcommand{\eep}{\mbox{$(e,e'p)$}}
\begin{document}
\pagestyle{prochead}


\title{Faddeev Approach to the Study of One- and Two-hole Spectral Functions}
\author{C. Barbieri}
  \email{carlob@hbar.wustl.edu}
  \homepage{http://www.physics.wustl.edu/~carlob}
  \affiliation
     {Department of Physics, Washington University,
         St.Louis, Missouri 63130, USA \\~\\  }

\author{W.~H.~Dickhoff}
  \email{wimd@wuphys.wustl.edu}
  \homepage{http://www.physics.wustl.edu/~wimd}
  \affiliation
    {Department of Physics, Washington University,
            St.Louis, Missouri 63130, USA \\~\\   }
  \affiliation
     {Department of Subatomic and Radiation Sciences, University of Gent,
           Proeftuinstraat 86, B-9000 Gent, Belgium \\~\\  }

\begin{abstract}
Theoretical calculations of one- and two-hole spectral functions  for
the \Oxg\ nucleus are still failing to describe some of the important
features observed experimentally.
Of critical importance for the solution of these issues is to obtain an
appropriate description of the interplay between hole-hole and particle-hole
excitations.   A formalism is reviewed here that allows the consistent
treatment of such phonons at a Random Phase approximation level.  Although
the  application of this formalism to \Oxg\ is still in the implementation
stage, some preliminary results are discussed here.
\end{abstract}
\maketitle
\setcounter{page}{1}

\section{Introduction}
\label{sec:introduction}

As a result of correlations between nucleons inside the nuclear medium,
the occupation probability of single-particle shells are lowered with respect to
the mean-field prediction. Experimentally, this can be observed as a quenching
of the absolute spectroscopic factors for the knockout of a nucleon from
a given shell.
Studies of \eep\ reactions have allowed the determination of spectroscopic
factors in many closed-shell nuclei~\cite{diep,sick,Lapik} demonstrating
that the removal probability for nucleons from these systems is reduced by
about 35\% with respect to the shell-model predictions.

For the \Oxg\ nucleus, the experimental spectroscopic strength~\cite{leus}
for the knockout of a proton from both the $p_{1/2}$ and $p_{3/2}$ shells
corresponds to about $60\%$. 
 The inclusion of relativistic effects
in the analysis of data is espected to change these results
by not more than a few~\%~\cite{udias}.
Theoretical calculations for this nucleus seem still far from being able
to reproduce
quantitatively the experiments and give numbers that depend on the
approximation scheme employed.
 Variational calculations, focussed on short-range correlations,
yield about a 10\% of strength removal~\cite{rad94,md94,Fabr01} before the
effects of the center of mass motion are included. Center of mass corrections
are known to raise the spectroscopic factor by about 7\%~\cite{O16jast}, thus
worsening the agreement with data. 
 An improvement of these results seems to require a treatment of
low-energy correlations with the inclusion two-hole--one-particle (2h1p)
states~\cite{FabrNM,fabrconf}.
This would be in line with the Green's function calculations of
Ref.~\cite{GeurtsO16}, in which 2h1p states were taken into account and
a reduction of about 25\% was obtained, still neglecting center of mass
corrections.
 A comparison with the experimental result (and their
uncertainties) suggests that about 20\% of the observed single-particle
strength removal remains unexplained.
Nevertheless, it is important to note that the results of Ref.~\cite{GeurtsO16}
indicate the low-energy correlations as an essential ingredient, needed for
a complete understanding of this puzzle.

 The merit of the calculation of Ref.~\cite{GeurtsO16} lies in the fact
that summing 2h1p states and their interactions to all orders, one can
achieve a simultaneous description of the effects of both 
particle-hole~(ph) and hole-hole~(hh) collective excitations,
including the interplay between them.
  These effects, though, were accounted for only at a Tamm-Dancoff
approximation (TDA) level.
In order to account for the coupling to collective excitations that are
actually observed in ${}^{16}{\rm O}$ it is necessary to at least consider 
a Random-Phase approximation (RPA) description of the isoscalar negative
parity states~\cite{czer}.
To account for the low-lying isoscalar positive parity states an even more
complicated treatment will be required.
Sizable collective effects are also present in the particle-particle (pp)
and hh excitations involving tensor correlations for isoscalar and
pair correlations for isovector states.

Taking a look at the experimental spectral function~\cite{leus},
one can also notice
that the total strength for proton emission of a $p_{3/2}$ proton
is experimentally
distributed over a main peak and two small fragments at slightly higher
missing energy.  This splitting is not
described by the calculations of Ref.~\cite{GeurtsO16} where only one pole
is generated that contains most of the single-particle strength.

The results of Ref.~\cite{GeurtsO16} have been used as a starting point
to study the \Oeepp\ reaction~\cite{geurts2h,geueepp}. 
In these works, the two-hole spectral function was obtained by coupling
two dressed single-particle (sp) propagators and then allowing
them to interact to all orders through a ladder type equation. 
   The distortaion of the two-hole overlap function caused by short-range 
effects was also included by adding defect functions associated with a
G-matrix interaction~\cite{CALGM} which was computed for the finite \Oxg \ 
nucleus.
These calculations
led to a proper description of the experimental cross section for 
two proton emission~\cite{ond1,ond2} and yield amplitudes for the
transition to the ground state of \Carb \ and to some of its lowest excited
states.

Nevertheless a completely satisfactory description is missing here also.
Recent developments of NN knockout reactions allow to distinguish several
states of the final nucleus and in particular it is now possible to disentangle
the two $2^+$  excited states of \Carb\  at 7.0 and 8.32~MeV~\cite{exp1,exp2}.
In spite of this, the theoretical calculations of Ref.~\cite{geurts2h} can
generate only one $2^+$ excited state.
It is interesting to note that the transition to both of the $2^+$ states can
be interpreted as the knockout of two protons from a $p_{1/2}$ and
a $p_{3/2}$ state.
Although this has not been directly investigated yet, it comes natural to
suppose that the fragmentation of the $p_{3/2}$ peak seen in the \Oeep \ 
reaction could play some role in the generation of the $2^+$
doublet in \Carb .
 Moreover, it must be noted that the strength of the $p_{1/2}$ and
$p_{3/2}$ peaks, generated in the calculations of the one-hole 
spectral function, enters directly in the calculation of the two-hole
amplitude.

It is therefore evident that one- and two-hole spectral functions are closely
related, the correct description of the first being relevant for a successful
description of the second and vice versa.  As a consequence, a formalism in
which both quantities are generated together in a consistent way is highly
desirable.
This can be achieved in the the framework of self-consistent Green's function
theory (SCGF) in which the equations of motion are expressed in terms
of fragmented propagators and are iterated until self-consistency is reached.

\begin{figure}[b]
 \begin{center}
    \parbox[b]{1.0\linewidth}{
      \includegraphics[width=1.\linewidth]{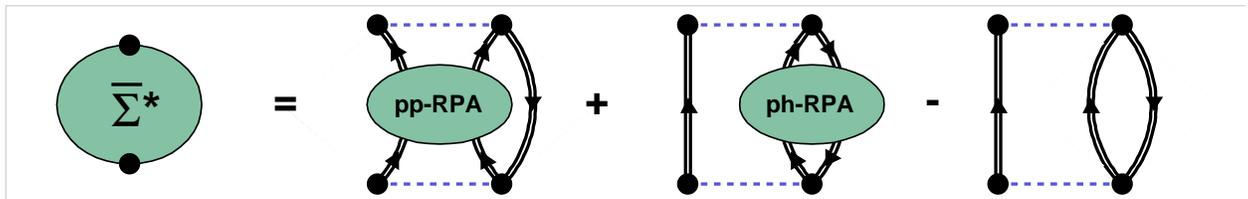}}
    \parbox[b]{7mm}{~}
    \parbox[b]{1.\linewidth}{
      \caption[]{\label{fig:pp_ph_sf}
        \small   Example of an approximation for the self-energy. Although
          this approximation contains both ph and pp correlations it would
          generate incorrect results, due to the need for subtracting the
          2nd order term to avoid double counting.}

      }
  \end{center}
\end{figure}

At the same time, it is also desirable to extend the calculations of
Ref.~\cite{GeurtsO16} to include the RPA description of pp~(hh) and ph phonons.
 This is not completely trivial. For example, the naive summation of diagrams
containing both pp and ph phonons leads to serious inconsistencies.
This approximation is depicted in Fig.~\ref{fig:pp_ph_sf}. 
The last of the three diagrams on the
right hand side is already contained in each of the other two and must
therefore be subtracted to avoid double counting.  
This subtraction introduces spurious poles in the Lehmann representation of the
self-energy and generates meaningless (not even normalizable) solutions
of the Dyson equation.
In addition, each of the first two terms in Fig.~\ref{fig:pp_ph_sf}
ignores the Pauli correlations
between the freely propagating line and the quasi-particles forming the phonons,
as noted in~\cite{Rijsdijk}.

\begin{figure}[ht]
 \begin{center}
    \parbox[b]{.35\linewidth}{
      \includegraphics[width=1.\linewidth]{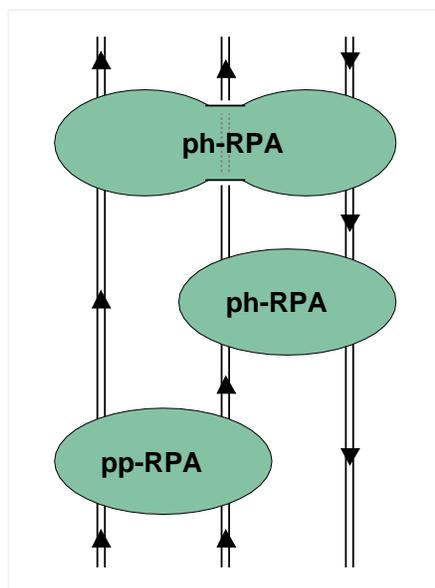}}
    \parbox[b]{7mm}{~}
    \parbox[b]{.4\linewidth}{
      \caption[]{\label{fig:faddallo}
         \small  The application of the Faddeev formalism allows to sum
                the effects of pp (hh) and ph RPA phonons to all orders,
                generating diagrams like this one. As a result of the
                all-orders summation, no inconsistent poles will
                appear in the self-energy.}

      }
  \end{center}
\end{figure}

 To achieve the inclusion of RPA effects, one needs to pursue a formalism
in which the contribution of the pp and ph phonons to the self-energy are
summed to all orders, like the example of Fig.~\ref{fig:faddallo}.
This, would avoid the subtraction of the second-order diagram.
Such a formalism has been proposed recently~\cite{barb} with the intent of
attacking the above issues. In this approximation scheme the effects of both 
pp (hh) and ph phonons can be evaluated at an RPA level and then
summed up to all orders by means of the Faddeev technique~\cite{Fadd1,glock}.
 The treatment of Pauli correlations is also improved over methods that
include ph RPA phonons directly in the self-energy,
since all exchange terms at the 2p1h level are consistently included.

An issue that one needs to be aware of in the application of the formalism
of Faddeev equations is the
apparence of spurious solutions. The Faddeev approach consists in substituting
the Sch\"odinger equations with a set of three coupled differential equations.
As a result, more solutions are generated and about 2/3 of them are spurious.
In the general three-body problem, though, the properties of such spurious
states are well known and they can be eliminated efficiently~\cite{Glockle99}.  
 The situation is a little more complex if the
Faddeev formalism is applied in a many-body context, with the intent
of studying the motion of quasi-particle and quasi-hole excitations.
 Here, in particular, the fulfillment of
closure relations for pp and ph
amplitudes is related to the behavior of the spurious Faddeev
eigenstates~\cite{barb}. This makes the choice of the interaction boxes to 
be used a critical problem, since without a proper treatment of this
relation the spurious solutions will irremediably mix with the
physically meaningful ones.

The Faddeev approach represents a step further with respect to the
calculations of Refs.~\cite{GeurtsO16,geurts2h} and it is our aim
to apply it to the study of one- and two-hole spectral functions of \Oxg. 
In this contribution, the basic formalism is
briefly reviewed.
 Some details and some technical issues
(as the elimination of spurious solutions) are described more 
completely in Ref.~\cite{barb} and will only be mentioned.
 The possible implementation of the SCGF scheme is also a key feature of the
present formalism and it will also be reviewed in some detail at the end of
Sec.~\ref{sec:SCGFT}

 The actual application to the nucleus of \Oxg\ is currently under way but
has not been completed and will be presented in a future paper. 
Nevertheless, some preliminary results have been generated for a simplified
model and they will be described in the last part, as an example.

\section{Faddeev formalism for 2h1p and 2p1h motion}

 The relevant quantities for the study of one- and two-nucleons knockout
reactions are the one- and two-hole spectral functions
\begin{equation}
S^{h}_\alpha (\omega) ~=~ \sum_k | {\mbox{$\langle {\Psi^{A-1}_k} \vert $}}
 c_\alpha {\mbox{$\vert {\Psi^A_0} \rangle$}} | ^2
 \; \delta ( \omega + E^{A-1}_k - E^A_0  ) \; ,
\label{eq:onehspf}
\end{equation}
\begin{equation}
S^{h}_{\alpha \beta} (\omega) ~=~  
    \sum_k | {\mbox{$\langle {\Psi^{A-2}_k} \vert $}}
 c_\alpha c_\beta {\mbox{$\vert {\Psi^A_0} \rangle$}} | ^2
 \;  \delta ( \omega + E^{A-2}_{k-} - E^A_0 ) \; .
\label{eq:twohspf}
\end{equation}
 
These can be obtained from the diagonal elements of the one- and two- body
Green's function respectively, which in their Lehmann representation take the
form~\cite{fetwa,AAA}
\begin{equation}
 g_{\alpha \beta}(\omega) ~=~ 
 \sum_n  \frac{ \left( {\cal X}^{n}_{\alpha} \right)^* \;{\cal X}^{n}_{\beta} }
                       {\omega - \varepsilon^{+}_n + i \eta }  ~+~
 \sum_k \frac{ {\cal Y}^{k}_{\alpha} \; \left( {\cal Y}^{k}_{\beta} \right)^*  }
                       {\omega - \varepsilon^{-}_k - i \eta } \; ,
\label{eq:fullg}
\end{equation}
and
\begin{equation}
  g^{II}_{\mu \nu , \alpha \beta}(\omega) ~=~ 
  \sum_{n+} \frac{\left( {\cal Z}^{n+}_{\mu \nu} \right)^* {\cal Z}^{n+}_{\alpha \beta}  }
              { \omega - \varepsilon^{\Gamma+}_{n+} + i \eta } ~-~
  \sum_{k-} \frac{{\cal Z}^{k-}_{\mu \nu} \left( {\cal Z}^{k-}_{\alpha \beta} \right)^*  }
              { \omega - \varepsilon^{\Gamma-}_{k-} - i \eta }  \; .
\label{eq:Gpp_leh}
\end{equation}
 In Eq.~(\ref{eq:fullg}),
${\cal X}^{n}_{\alpha} = {\mbox{$\langle {\Psi^{A+1}_n} \vert $}}
 c^{\dag}_\alpha {\mbox{$\vert {\Psi^A_0} \rangle$}}$%
~(${\cal Y}^{k}_{\alpha} = {\mbox{$\langle {\Psi^{A-1}_k} \vert $}}
 c_\alpha {\mbox{$\vert {\Psi^A_0} \rangle$}}$) are the
spectroscopic amplitudes for the excited states of a system with
$A+1$~($A-1$) particles and the poles $\varepsilon^{+}_n = E^{A+1}_n - E^A_0$%
~($\varepsilon^{-}_k = E^A_0 - E^{A-1}_k$) correspond to the excitation
energies with respect to the $A$-body ground state.  Completely analogous
definitions are employed in Eq.~(\ref{eq:Gpp_leh}) for the transition
amplitudes and excitation energies related to the states of a system
with $A+2$~($A-2$) particles.

  In the nuclear case, a strong coupling exists between the sp degrees of
freedom and both collective low-lying states as well as high-lying states.
The latter coupling is related to the strong short-range repulsion
in the nuclear force.
 The resulting fragmentation
of the sp strength (as observed in experimental data) suggests that this
feature must already be included in the description of these couplings. This
should be done by expressing the equations used to evaluate the propagators
$g_{\alpha \beta}(\omega)$ and $g^{II}_{\mu \nu , \alpha \beta}(\omega)$
in terms of the solution $g_{\alpha \beta}(\omega)$ itself.
This self-consistency feature also emerges in an exact formulation, involving 
the coupling to two-, three- and $A$-body propagators, which can be
derived using the equation of motion method~\cite{masch}. 

 Keeping in mind this approach, we compute $g_{\alpha \beta}(\omega)$
as a solution of the Dyson equation
\begin{equation}
g_{\alpha \beta}(\omega) ~=~ g^{(0)}_{\alpha \beta}(\omega) ~+~ 
                              g^{(0)}_{\alpha \gamma}(\omega)
                              \Sigma^{\star}_{\gamma \delta}(\omega)
                              g_{\delta \beta}(\omega)  \; ,
\label{eq:Dyson}
\end{equation}
where $\Sigma^{\star}_{\alpha \beta}(\omega)$ is the irreducible self-energy.
Here and in the following, sums over repeated indices are implied.

By considering the equation of motion for $g_{\alpha \beta}(\omega)$, one
obtains that
$\Sigma^{\star}_{\alpha \beta}(\omega)$ can be written as the sum of two terms
\begin{equation}
 \Sigma^{\star}_{\alpha \beta}(\omega) ~=~  \Sigma^{HF}_{\alpha \beta} 
   ~+~ \frac{1}{4} \, V_{\alpha \lambda , \mu \nu} ~
     R_{\mu \nu \lambda , \gamma \delta \varepsilon}(\omega)
         ~ V_{\gamma \delta , \beta \varepsilon}  \; ,
\label{eq:Sigma1}
\end{equation}
where $\Sigma^{HF}_{\alpha \beta}$ represents the
Hartree-Fock part of the self-energy, which can be computed straightforwardly
from the solution $g_{\alpha \beta}(\omega)$ itself.
 The $V_{\alpha \lambda , \mu \nu}$ represent the antisymmetrized matrix
elements of the residual interaction.

The non trivial element to be inserted in the calculation is the propagator
$R_{\mu \nu \lambda , \gamma \delta \varepsilon}(\omega)$, appearing in the
last term of Eq.~(\ref{eq:Sigma1}), which describe the
motion of excitations consisting of at least two-particle--one-hole (2p1h)
or 2h1p states.
 It contains the sum of all six points diagrams that are one-particle
irreducible, i.e. that cannot be separated by cutting a single line.

 In general, $R_{\mu \nu \lambda , \gamma \delta \varepsilon}(\omega)$ is the
solution of a Bethe-Salpeter type equation involving propagators depending
on four times (i.e. three energies) as well as four- and six- point
kernels~\cite{Win72}.
This is appears too complex to be solved numerically and one needs
to find an appropriate approximation for the propagator
$R_{\mu \nu \lambda , \gamma \delta \varepsilon}(\omega)$.
 The formalism proposed in Ref.~\cite{barb} aims to compute this object
keeping the effects of RPA phonons but at the same time restricting oneself
to equations involving only two-times propagators, thus producing a set of
equations that one can deal with. This is achieved by applying the Faddeev
equations technique.

 Also, the resulting formalism splits up in two separate expansions for
the 2p1h and the 2h1p components. Although the hole spectral function is
of primary interest for comparison with  experimental data, it must be
stressed that both 2p1h and 2h1p components are needed to generate the
self-consistent solution for the sp propagator.
Since the formalism involved is the same for both components, we will
describe only the forward-going (2p1h) expansion in the rest of this
section. The equations for the 2h1p case are completely analogous.

\subsection{Faddeev equations}

As a convention, we assume that the first two lines propagating through the
expansion of $R_{\mu \nu \lambda , \gamma \delta \varepsilon}(\omega)$
represent particle excitations while the last represents a hole, as
in Fig.~\ref{fig:faddallo}. Following 
standard notation in the literature~\cite{JoachBk}, the i-th Faddeev component
$R^{(i)}_{\mu \nu \lambda , \alpha \beta \gamma}$ will represent 
the sum of all diagrams ending with an interaction between legs $j$
and $k$, with $(i,j,k)$ cyclic permutations of $(1,2,3)$.
 Obviously, the component $R^{(3)}$ refer to the interactions in the pp 
channel, while the two components $R^{(1)}$ and $R^{(2)}$ will refer to
the interaction between a particle and a hole.
The last two are trivially related to each other 
by  the exchange of the first two indices.

 Also, for reasons to be mentioned in the next subsection, the $R^{(i)}$
need to be redefined in such a way that their matrix
elements also depend on the indices ($n$, $n'$, $k$), which label the
fragments of the external propagator lines (compare with the definitions
below Eq.~(\ref{eq:fullg})~).
 This implies that the eigenvalue equations will involve
summations on both the sp indices ($\alpha$, $\beta$, $\gamma$) and the
ones corresponding to the fragmentation, ($n_\alpha$, $n_\beta$, $k_\gamma$).
 The 2p1h propagator and its Faddeev components are recovered only at
the end of calculations by summing the solutions over all values
of ($n_\alpha$, $n_\beta$, $k_\gamma$) and~($n_\mu$,~$n_\nu$,~$k_\lambda$).

 Putting together all the above ingredients, the resulting approximation to
the Faddeev equations can be rewritten in a 
way where all the propagators involved depend only on one energy variable
(or two time variables).
 The 2p1h part of this expansion can be written as follows
\begin{eqnarray}
  \lefteqn{
  R^{(i)}_{\mu    n_\mu    \nu   n_\nu   \lambda k_\lambda , 
           \alpha n_\alpha \beta n_\beta \gamma  k_\gamma}(\omega) }
        \hspace{.3in} & &
\nonumber  \\
   &=&  \frac{1}{2}\left( 
      {G^0}^>_{\mu    n_\mu    \nu   n_\nu   \lambda k_\lambda , 
               \alpha n_\alpha \beta n_\beta \gamma  k_\gamma}(\omega)
    - {G^0}^>_{\nu   n_\nu   \mu    n_\mu    \lambda k_\lambda ,
               \alpha n_\alpha \beta n_\beta \gamma  k_\gamma}(\omega) \right)
\nonumber  \\
  & & +~ {G^0}^>_{\nu  n_\nu   \mu  n_\mu   \lambda  k_\lambda ,
                  \mu' n_\mu'  \nu' n_\nu'  \lambda' k_\lambda'}(\omega) ~
    \Gamma^{(i)}_{\nu'  n_\nu'  \mu'  n_\mu'   \lambda'  k_\lambda' ,
                  \mu'' n_\mu'' \nu'' n_\nu''  \lambda'' k_\lambda''}(\omega) ~
\nonumber  \\
  & & ~ ~ ~ \times ~
  \left( R^{(j)}_{\mu''    n_\mu''    \nu''   n_\nu ''  \lambda'' k_\lambda'' , 
                  \alpha n_\alpha \beta n_\beta \gamma  k_\gamma}(\omega) ~+~
         R^{(k)}_{\mu''    n_\mu''    \nu''   n_\nu ''  \lambda'' k_\lambda'' , 
                  \alpha n_\alpha \beta n_\beta \gamma  k_\gamma}(\omega)
       \right)  \; ,  ~ i=1,2,3 ~ .
\label{eq:FaddTDA}
\end{eqnarray}
 In Eq.~(\ref{eq:FaddTDA}), ${G^0}^>$ is the forward-going part of the 2p1h
propagator for three dressed but noninteracting lines.
 Using the notations introduced after Eq.~(\ref{eq:fullg}) we have
\begin{equation}
  {G^0}^>_{\mu    n_\mu    \nu   n_\nu   \lambda k_\lambda , 
           \alpha n_\alpha \beta n_\beta \gamma  k_\gamma}(\omega) ~=~
        \delta_{n_\mu , n_\alpha} \;
        \delta_{n_\nu , n_\beta} \;
        \delta_{k_\lambda , k_\gamma} ~
     \frac{ \left( 
         {\cal X}^{n_\mu}_{\mu}
         {\cal X}^{n_\nu}_{\nu}
         {\cal Y}^{k_\lambda}_{\lambda}
            \right)^* \; 
         {\cal X}^{n_\alpha}_{\alpha}
         {\cal X}^{n_\beta}_{\beta}
         {\cal Y}^{k_\gamma}_{\gamma} }
    { \omega - ( \varepsilon^+_{n_\alpha} + \varepsilon^+_{n_\beta} -
                                     \varepsilon^-_{k\gamma} ) + i \eta } \; .
\label{eq:G0fw}
\end{equation}

The Faddeev vertices
$\Gamma^{(i)}_{\nu'  n_\nu'  \mu'  n_\mu'   \lambda'  k_\lambda' ,
\mu'' n_\mu'' \nu'' n_\nu''  \lambda'' k_\lambda''}(\omega)$  contain the
collective excitations in the pp and ph channels and can be evaluated in RPA,
as explained in the next subsection.

 Once the Faddeev equations are solved and the usual representation of the
Faddeev components $R^{(i)}_{\mu \nu \lambda , \alpha \beta \gamma}$ is
recovered by summing over the indices
($n_\alpha$,~$n_\beta$,~$k_\gamma$) and~($n_\mu$,~$n_\nu$,~$k_\lambda$),
the complete 2h1p propagator can be obtained as the sum of all the components:
\begin{equation}
    R_{\mu \nu \lambda , \alpha \beta \gamma}(\omega) ~=~
   \sum_{i=1,2,3}     R^{(i)}_{\mu \nu \lambda , \alpha \beta \gamma}(\omega)
     ~-~ \frac{1}{2} \left( {G^0}^>_{\mu \nu \lambda , \alpha \beta \gamma}(\omega)
        - {G^0}^>_{\nu \mu \lambda , \alpha \beta \gamma}(\omega) \right) \; ,
\label{eq:faddfullR}
\end{equation}
The free forward-going  propagator ${G^0}$ was
already included in the Eqs.~(\ref{eq:FaddTDA}) and therefore is to
be subtracted in Eq.~(\ref{eq:faddfullR}) to avoid double countings.
 It must be stressed though that this subtraction does not generate any
inconsistencies in the self-energy, as it was the case for the approximation
of Fig.~\ref{fig:pp_ph_sf}. Rather, this subtraction is required to
exactly cancel similar, troublesome, terms that are included in the single
Faddeev components.

\subsection{Faddeev vertices}

 The interaction matrices $\Gamma^{(pp)}_{\mu \nu , \alpha \beta}(\omega)$
for the pp channel and $\Pi^{(ph)}_{\mu \nu , \alpha \beta}(\omega)$
for the ph channel, are obtained by solving the ladder equation in the
Dressed RPA (DRPA) scheme, which employs dressed propagators as they are
obtained by solving the Dyson equation~(\ref{eq:Dyson}).

\begin{figure}[ht]
 \begin{center}
    \parbox[b]{0.75\linewidth}{
      \includegraphics[width=0.72\linewidth]{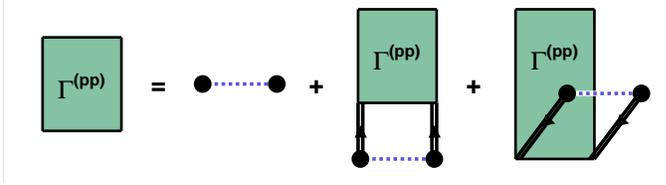}}
    \parbox[b]{0.22\linewidth}{
      \caption[pp DRPA]{\label{fig:ppDRPA}
        \small  DRPA equation for the $\Gamma^{(pp)}$ matrix. }
      }
  \end{center}
\end{figure}

 For the pp case, the DRPA equation is graphically depicted in
Fig.~\ref{fig:ppDRPA}.
 The gamma matrix $\Gamma^{(pp)}_{\mu \nu , \alpha \beta}(\omega)$ obtained
in this way is
the generalization of the two-body T-matrix for two nucleons and
represents our approximation for the two-particle propagator of
Eq.~(\ref{eq:Gpp_leh}), when  the external legs are amputated.
   Completely analogous considerations apply to the ph interaction matrix
$\Pi^{(ph)}_{\mu \nu , \alpha \beta}(\omega)$.
 
 The RPA phonons generated in this way depend on dressed propagators 
and can be included in the Faddeev formalism outlined above.
  It is worth noting that discarding the last diagram on the right hand side
of Fig.~\ref{fig:ppDRPA}, and in the corresponding ph counterpart,
the whole formalism reduces to the TDA expansion of Ref.~\cite{GeurtsO16}.

\begin{figure}[ht]
 \begin{center}
    \parbox[b]{0.6\linewidth}{
      \includegraphics[width=1.\linewidth]{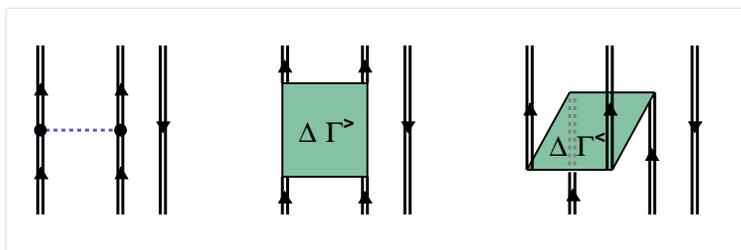}}
    \parbox[b]{1.\linewidth}{
      \caption[pp DRPA]{\label{fig:Gammai_vs_DRPAS}
     \small Here $\Delta \Gamma^>$ and $\Delta \Gamma^<$ are the forward- and
        backward-going part of the energy dependent contribution to the 
        pp DRPA vertex~(\ref{eq:Gpp_leh}).
         The contribution of these three diagrams can be factorized in an 
        expression of the form ${G^0}^> \; \Gamma^{(3)} \; {G^0}^>$ 
        only after having redefined the propagators ${G^0}^>$ and 
        $\Gamma^{(3)}$ to depend also on the particle and hole
        fragmentation indices ($n$,$n'$,$k$).
         The last diagram has a smaller effect on the physical solutions
        of the problem, although it is essential for the elimination of
        spurious solutions.  }
      }
  \end{center}
\end{figure}

 The vertices to be employed in the Faddeev
equations~(\ref{eq:FaddTDA}) will consists of the interaction boxes
just described to which a freely propagating dressed line is added,
as displayed in Fig.~\ref{fig:Gammai_vs_DRPAS}.
 The exact formula for the pp channel is as follows:
\begin{eqnarray}
\lefteqn{
  \Gamma^{(3)}_{\mu n_\mu \nu n_\nu \lambda k_\lambda ,
           \alpha n_\alpha \beta n_\beta \gamma k_\gamma}(\omega) ~=~
        {\textstyle {1\over 2}} \frac{\delta_{k_\lambda , k_\gamma}}
                 { \sum_\sigma \left| {\cal Y}^{k_\lambda}_\sigma \right|^2 } }
    \hspace{.5in} & &
\nonumber \\
 ~ ~ ~ ~ ~ ~ ~ ~ ~ ~ ~ ~
 & & \times ~ \left\{      V_{\mu \nu , \alpha \beta}  ~+~
       \sum_{n+} 
         \frac{\left( \Delta^{n+}_{\mu \nu} \right)^* \Delta^{n+}_{\alpha \beta}  }
              { \omega - ( \varepsilon^{\Gamma+}_{n+} - \varepsilon^-_{k_\lambda} ) + i \eta } \right.
\nonumber \\
 & & ~+~ \left.   \sum_{k-} 
         \frac{ [ \omega - \varepsilon^+_{n_\mu} - \varepsilon^+_{n_\nu}
                          - \varepsilon^+_{n_\alpha} - \varepsilon^+_{n_\beta}
                           + \varepsilon^-_{k_\lambda} + \varepsilon^{\Gamma-}_{k-} ]
        \;  \Delta^{k-}_{\mu \nu} \left( \Delta^{k-}_{\alpha \beta} \right)^*  }
              { (\varepsilon^{\Gamma-}_{k-} - \varepsilon^+_{n_\mu} - \varepsilon^+_{n_\nu} )
                (\varepsilon^{\Gamma-}_{k-} - \varepsilon^+_{n_\alpha} - \varepsilon^+_{n_\beta} ) }
 \right\} \; ,
\label{eq:Gamma3}
\end{eqnarray}
were the $\Delta^{n+}$~($\Delta^{k-}$) amplitudes are computed through the
DRPA equation of Fig.~\ref{fig:ppDRPA} and are related to the amplitudes of
Eq.~(\ref{eq:Gpp_leh}) by
$\Delta^{n+}_{\mu \nu} = V_{\mu \nu , \alpha \beta} \, {\cal Z}^{n+}_{\alpha \beta}$
($\Delta^{k-}_{\mu \nu} = {\cal Z}^{k-}_{\alpha \beta} \,
V_{\alpha \beta , \mu \nu} $).

The complication of defining the interaction vertex $\Gamma^{(3)}$ not only
in terms of the sp indices but also in terms of the respective fragmentations
($n_\mu$,~$n_\nu$,~$k_\lambda$) and~($n_\alpha$, $n_\beta$, $k_\gamma$) is a
consequence of the two main requirements imposed: that is, $\Gamma^{(3)}$
has to depend on only two times and the free line added in
Fig.~\ref{fig:Gammai_vs_DRPAS} must be a dressed propagator.
 As it can be seen from
Eq.~(\ref{eq:Gamma3}),  the energy poles $\varepsilon^-_{k_\lambda}$,
relative to the fragments of this line,
appear in the equation mixed with the corresponding poles
of the $\Gamma^{(pp)}_{\mu \nu , \alpha \beta}(\omega)$'s phonons.
This situation couldn't be handled without the above
prescription~\cite{DRPApaper,barb}.
 Instead, with this procedure it is still possible to write an expression
for the diagrams of Fig.~\ref{fig:Gammai_vs_DRPAS} in the form
${G^0}^>(\omega) \; \Gamma^{(3)}(\omega) \; {G^0}^>(\omega)$ or, equivalently,
to write the Faddeev equations~(\ref{eq:FaddTDA}).

As a last remark, we note that the last term on the right hand side
of Eq.~(\ref{eq:Gamma3}) comes
from the last diagram of Fig.~\ref{fig:Gammai_vs_DRPAS}.  This term contains
large energy denominators and it is expected to give relatively small
contributions.  Nevertheless, it can be proved that it is essential to
include it in order to impose the appropriate behavior
of the spurious solutions generated by the Faddeev equations
(and eventually to discard them)~\cite{barb}.  Thus also this diagram needs
to be kept.

\subsection{Self-consistent Scheme}
\label{sec:SCGFT}

 The set of Faddeev equations~(\ref{eq:FaddTDA}), together with the
expressions~(\ref{eq:Dyson}) and~(\ref{eq:Sigma1}),
furnish a scheme in which  one- and two-body
spectral functions can be evaluated in terms of an already dressed
single-particle Green's function.
  This allows the possibility of including,
already from the start, the features of the single-particle motion
that go beyond a mean-field description.  Once such features are
included in  the input propagator, their effects
on the one- and two-body motion will be automatically included in the
calculations.
It is important to observe that the expansion of the irreducible propagator
$R_{\mu \nu \lambda , \gamma \delta \varepsilon}(\omega)$
in Eq.~(\ref{eq:Sigma1}),
derived employing the equation of motion, is given in terms of the real
 one-body propagator.   Thus, in principle, it is the exact solution of the 
many-body problem that is supposed to be employed in the Faddeev calculations
described above. 

\begin{figure}[b]
 \begin{center}
    \parbox[b]{0.8\linewidth}{
      \includegraphics[width=1.\linewidth]{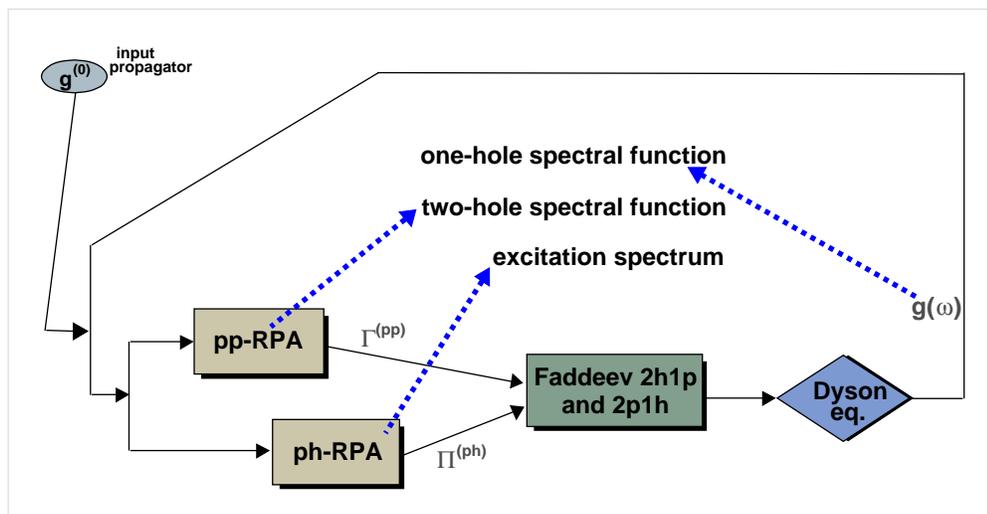}}
    \parbox[b]{1.\linewidth}{
      \caption[pp DRPA]{\label{fig:SCGF}
   \small   Iteration scheme used to reach self-consistency between the input
        propagator and the result of  the Dyson equation. At each iteration
        new approximations for the one- and two- body
        propagators~(\ref{eq:fullg}) and~(\ref{eq:Gpp_leh}) are obtained.   }
      }
  \end{center}
\end{figure}

 The approach of self-consistent Green's function theory consists in starting 
with an approximation for the input Green's function (usually an Hartee-Fock
propagator or other independent particle model). From this, the pp (hh), the
ph and the Faddeev propagators can be evaluated with the formalism described.
Then, the solution of the Dyson Eq.~(\ref{eq:Dyson}) will give a better 
approximation
of the s.p. propagator, that can be employed in the second calculation.
 As Shown in Fig.~\ref{fig:SCGF}, the whole procedure is iterated many
times until consistency is found between two successive solutions (i.e.
between the input propagator and the solution that it generates).

 The attractive feature of this procedure is not only restricted to the fact
that the effects of fragmentation are included in the calculations. But
also, expressions for the one- and  two-hole motion and  the ph phonons
(which give information on the excitation spectrum of the system) are
generated all at the same time, the relative effects of each one on the
others being taken into account.

  Obviously, the exact form of the single-particle propagator is very complex,
containing many poles and a continuum. Such richness of details
cannot be easily included in the above calculation.
 In our case, though,  we are interested in the low energy
behavior of the single-particle motion and the few poles that contain the
main sp strength are expected to be much more relevant than the others. 
  One possible prescription to limit the number of poles
in the input propagator at each iteration
consists in keeping only the most significant peaks and collapsing
all the others in an effective pole.
  This procedure was already employed in applying the SCGF method to the
study of pairing in superfluid nuclei~\cite{jyuan,wim99}.

 The effects coming from short-range correlations, can be taken into account
by substituting the bare NN potential $V_{\alpha \beta , \gamma \delta}$
with a G-matrix interaction.
  In the same way,  the effects of short-range interaction can be included in
the two-hole spectral function  by adding the relative defect functions
as done in Ref.~\cite{geurts2h,geueepp}.

\section{Quadrupole-plus-Pairing Model}

 The implementation of the Faddeev RPA formalism available at the moment
allows us to make a test with a simplified model for the \Oxg \ nucleus.
 We considered a model space consisting of harmonic oscillator wave functions
up to four major closed shells ($2p$ and $1f$ states) plus a $g_{9/2}$ single
particle state.
 Within this space, we choose a so called ``pairing-plus-quadrupole''
Hamiltonian~\cite{schuckbook}, in which one a body potential $\hat{U}$ is
employed  to generate the correct single-particle energies and the residual
interaction in composed of a quadrupole-quadrupole $\hat{V}_{\cal Q}$ and a 
pairing $\hat{V}_{\cal P}$ term:
 \begin{eqnarray}
\hat{H} ~&=&~\hat{U} ~+~ \hat{V}_{\cal Q} ~+~ \hat{V}_{\cal P} 
  \nonumber \\  
       &=&~ \sum_{\alpha} \varepsilon^0_\alpha ~c^{\dag}_\alpha c_\alpha 
       ~+~ g_{_{\cal Q}} ~  : \hat{Q}^{\dag} \hat{Q} :
       ~+~  g_{_{\cal P}} ~  \hat{P}^{\dag} \hat{P}  \; ,
\label{eq:Hamiltonian}
\end{eqnarray}
 where
$\hat{P}~\equiv \sum_{\alpha} c_\alpha c_{\bar{\alpha}}$ and
$\hat{Q}~\equiv \sum_{\alpha , \beta}
{\mbox{$\langle {\alpha} \vert $}} r^2 Y_{2,m}(\Omega_r)
{\mbox{$\vert {\beta} \rangle$}} c^{\dag}_\alpha c_{\beta}$
are the pairing and the quadrupole operators,  the symbol ``$:$'' represents
the normal ordering and $\bar{\alpha}$ is the time reversal of 
state~$\alpha$.  An oscillator parameter of $b~=~1.76~fm$ was used in
computing the quadrupole matrix elements.
 Both $\hat{V}_{\cal P}$ and $\hat{V}_{\cal Q}$ are independent of the spin
and isospin degrees of freedom.

 The values of $\varepsilon^0_\alpha$ for all the sp states were computed in 
Ref.~\cite{GeurtsO16} in a Brueckner-Hartree-Fock calculation, based on a
G-matrix interaction~\cite{CALGM} derived from the Bonn-C
potential~\cite{bonnc}.
 The $\varepsilon^0_\alpha$ used here are are taken from that reference
but with some slight modification of the levels close to the Fermi energy,
in order to roughly reproduce the experimental missing energies
in the Faddeev RPA calculation. The exact values employed here are given in
table~\ref{tab:param}.
\begin{table}[t]
 \begin{center}
    \parbox[t]{1.\linewidth}{
      \caption[]{\label{tab:param}
         \small  Single particle energies $\varepsilon^0_\alpha$ and
         interaction strengths used in Eq.~(\ref{eq:Hamiltonian}).
          The sp energies were computeded from Ref.~\cite{GeurtsO16}, while
           $g_{_{\cal Q}}$ and $g_{_{\cal P}}$ were fitted to roughly reproduce 
           the excitation energy of the $3^-$ ph phonon.   }
      }
 \begin{tabular}{ccccccccccccc}
  \hline \hline
  \\
 $\alpha$ & $\;$ & $\; \; 1s_{1/2} \; \; $ & $ \; \; 1p_{1/2} \; \; $ & $ \; \; 1p_{3/2} \; \; $ & $ \; \; 1d_{5/2} \; \; $ & $ \; \; 2s_{1/2} \; \; $ & $ \; \; 1d_{3/2} \; \; $ & $ \; \; 1f_{7/2} \; \; $ & $ \; \; 2p_{3/2} \; \; $ & $ \; \; 1f_{5/2} \; \; $ & $ \; \; 2p_{1/2} \; \; $ & $ \; \; 1g_{9/2} \; \; $  \\
  \\
  \hline
  \\
$\varepsilon^0_\alpha (MeV)$ &  & $-35.0$ & $-18.2$ & $-9.3$ & $-1.8$ & $-0.1$ & $4.4$ & $17.4$ & $16.0$ & $23.5$ & $17.7$ & $30.0$  \\
  \\
  \hline \hline
  \\
  & & & \multicolumn{4}{c}{$g_{_{\cal Q}} = -1.14 \; MeV/fm^4$} & & \multicolumn{4}{c}{$g_{_{\cal P}} = +15.0 \; MeV$} &   \\
  \\
  \hline \hline
 \end{tabular}
  \end{center}
\end{table}

 The remaining two parameters,  $g_{_{\cal Q}}$
and $g_{_{\cal P}}$, were fitted to other experimental energies. In particular,
$g_{_{\cal Q}}$ was chosen to roughly reproduce the $3^-$ phonon of \Oxg ,
while $g_{_{\cal P}}$ was fitted aiming to reproduce the experimental missing 
energy of
the \Carb \ ground state. The latter was found to be not very sensitive and was
then fixed to the value given in table~\ref{tab:param}.

\begin{table}[t]
 \begin{center}
    \parbox[t]{1.\linewidth}{
      \caption[]{\label{tab:results}
       \small   Results for  hole spectroscopic factors of the main
       $p_{1/2}$ and $p_{3/2}$ peaks, obtained with
         Hamiltonian~(\ref{eq:Hamiltonian}). }
      }
 \begin{tabular}{ccccccc}
  \hline \hline
  \\
  $\qquad$  & & &  TDA  & & RPA & $\qquad$ \\
  \\
  \hline 
  \\
 $\qquad$  & $p_{1/2}$ & $\qquad$  &  .632  & $\qquad$  &  .588 & $\qquad$ \\
  \\
 $\qquad$  & $p_{3/2}$ & $\qquad$  &  .668  & $\qquad$  &  .647 & $\qquad$ \\
  \\
  \hline \hline
 \end{tabular}
  \end{center}
\end{table}

  Using the Hamiltonian~(\ref{eq:Hamiltonian}), the self-energy was computed
employing the Faddeev
formalism with pp (hh) and ph phonons evaluated in both TDA and RPA
schemes.
  As already noticed, the TDA approximation was already employed in the 
calculation of Ref.~\cite{GeurtsO16}.

 As seen from table~\ref{tab:results}, the results obtained
for the $p_{1/2}$ and $p_{3/2}$ spectroscopic factors are 4-5~\% smaller in
the RPA case and seem to go in the expected way.

 It's worth to recall that the results from this model have been obtained
within a single iteration, without the effects of short-range correlations
and with a oversimplified  interaction. Thus the values obtained for the
absolute spectroscopic factors are not really meaningful and are expected
to change significantly in a realistic calculation.
 Self-consistency can play a relevant role too.

 Nevertheless, a comparison between the results obtained in TDA and RPA
calculation seems reasonable and can give some idea of the relative
importance of RPA effects.

\section{Summary}

The present theoretical description of the distribution of spectroscopic strength
at low energies lacks important ingredients for a successful
comparison with experimental data.
One of these ingredients is a proper description of the coupling
of sp motion to low-lying collective modes that are present in the system.
Recent calculations for ${}^{16}{\rm O}$~\cite{GeurtsO16}, for example,
only include a TDA description of these collective modes.
 This contribution outlines a new method that has been proposed
recently~\cite{barb} that aims to study the influence of pp and ph RPA
correlations on the sp propagator for a system with a finite
number of fermions.
 This method is formulated in the context of SCGF theory 
by evaluating the nucleon self-energy in terms of
the 2p1h and 2h1p propagators.
 The description of the 2p1h (or 2h1p) excitations has been studied by using the
Faddeev formalism, which is usually applied to solve the three-body problem.
 As mentioned in the text, the detailed choice of the form of the 
interaction matrices as well as the formulation of the problem are imposed
by practical issues, like the existence of spurious solutions.

 It should be noted that the Faddeev approach outlined here, has also the
advantage that
it can be naturally extended to the inclusion of more complicated excitations
like the extended DRPA~\cite{DRPApaper}.

 The application of this formalism is underway for the nucleus of \Oxg.
 Although no physically meaningful results have been obtained yet, the
Faddeev~RPA equations have been applied to a simple model,  demonstrating the
feasibility of these calculations.
 The results obtained with this model yield somewhat stronger quenching of
the $p_{1/2}$ and $p_{3/2}$ spectroscopic factors when RPA correlations
are included, thus going in the expected direction.

 A realistic self-consistent calculation in which a G-matrix derived
from realistic interaction is employed is underway and the results 
will be reported elsewhere.

\section*{Acknowledgments}
This work is supported by the U.S. National Science Foundation under
Grant No. PHY-9900713.


\end{document}